\documentclass[fleqn,usenatbib]{mnras}

\usepackage{newtxtext,newtxmath}
\usepackage[T1]{fontenc}
\usepackage{multirow}

\usepackage{graphicx}
\usepackage{amsmath}
\usepackage{orcidlink}

\newcommand{\schro}{Schrödinger}
\newcommand{\xb}{\mathbf{x}}
\newcommand{\eV}{\mathrm{eV}}
\newcommand{\kpc}{\mathrm{kpc}}
\newcommand{\Msun}{{\rm M}_\odot}

\usepackage{xcolor}
\usepackage[normalem]{ulem}

\title{Nested solitons in two-field fuzzy dark matter}

\author[H. N. Luu et al.]{Hoang Nhan Luu\,\orcidlink{0000-0001-9483-1099}$^{1,2,8}$\thanks{E-mail: hnluu@connect.ust.hk}, Philip Mocz\,\orcidlink{0000-0001-6631-2566}$^{3}$, Mark Vogelsberger\,\orcidlink{0000-0001-8593-7692}$^{2}$, Simon May\,\orcidlink{0000-0002-2781-6304}$^{4,5}$, Josh Borrow\,\orcidlink{0000-0002-1327-1921}$^{2}$,
\newauthor
S.-H. Henry Tye\,\orcidlink{0000-0002-4386-0102}$^{1,6}$, Tom Broadhurst\,\orcidlink{0000-0002-8785-8979}$^{7,8,9}$
\\
$^{1}$Department of Physics and Jockey Club Institute for Advanced Study, The Hong Kong University of Science and Technology, Hong Kong\\
$^{2}$Department of Physics, Kavli Institute for Astrophysics and Space Research, Massachusetts Institute of Technology, Cambridge, MA 02139, USA \\
$^{3}$Lawrence Livermore National Laboratory, 7000 East Ave, Livermore, CA 94550, USA\\
$^{4}$Perimeter Institute for Theoretical Physics, 31 Caroline Street North, Waterloo, ON, N2L 2Y5, Canada\\
$^{5}$Department of Physics, North Carolina State University, Raleigh, NC, 27695-8202, USA\\
$^{6}$Department of Physics, Cornell University, Ithaca, NY 14853, USA \\
$^{7}$University of the Basque Country UPV/EHU, Department of Theoretical Physics, Bilbao, E-48080, Spain \\
$^{8}$Donostia International Physics Center, Basque Country UPV/EHU, San Sebastian, E-48080, Spain \\
$^{9}$Ikerbasque, Basque Foundation for Science, Bilbao, E-48011, Spain
}

\date{}

\pubyear{2023}

\begin{document}
\label{firstpage}
\pagerange{\pageref{firstpage}--\pageref{lastpage}}
\maketitle

\begin{abstract}
Dark matter as scalar particles consisting of multiple species is well motivated in string theory where axion fields are ubiquitous. A two-field fuzzy dark matter~(FDM) model features two species of ultralight axion particles with different masses, $m_1 \neq m_2$, which is extended from the standard one-field model with $m_a \sim 10^{-22} \,{\rm eV}$. Here we perform numerical simulations to explore the properties of two-field FDM haloes. We find that the central soliton has a nested structure when $m_2 \gg m_1$, which is distinguishable from the generic flat-core soliton in one-field haloes. However, the formation of this nested soliton is subject to many factors, including the density fraction and mass ratio of the two fields. Finally, we study non-linear structure formation in two-field cosmological simulations with self-consistent initial conditions and find that the small-scale structure in two-field cosmology is also distinct from the one-field model in terms of DM halo counts and soliton formation time.
\end{abstract}


%%%%%%%%%%%%%%%%%%%%%%%%%%%%%%%%%%%%%%%%%%%%%%%%%%

%%%%%%%%%%%%%%%%% BODY OF PAPER %%%%%%%%%%%%%%%%%%

\section{Introduction}

The nature of dark matter (DM) remains elusive since its first evidence was found almost a century ago by Fritz Zwicky~\citep{Zwicky:1937}. The cold dark matter~(CDM) hypothesis, where DM is made of non-relativistic, collision-less particles, has successfully explained the anisotropies of the Cosmic Microwave Background and the large-scale structure of the universe; see \cite{Planck:2018vyg} and \cite{eBOSS:2020yzd} for the most recent observational data from Planck and eBOSS, see \cite{Vogelsberger:2019ynw} for a review on state-of-the-art cosmological simulations of galaxy formation. From a microscopic perspective, the models of weakly interacting massive particles (WIMPs), where DM particles are presumably heavy and have interactions of electroweak scales with Standard Model particles, have been the leading candidates for CDM in the past decades. However, CDM is also facing several challenges due to inconsistencies between $N$-body simulations and observations on small scales, including the ``core--cusp problem''~\citep{deBlok:2009}, the ``too-big-to-fail problem''~\citep{Garrison-Kimmel:2014vqa, Boylan-Kolchin:2011}, the diversity problem of rotation curves~\citep{Oman:2015xda} -- together with the null results of WIMP signals in many laboratory searches~\citep{Roszkowski:2017nbc}.

Even though baryonic feedback, e.\,g. from supernova explosions or stellar winds~\citep{Governato:2012, DiCintio:2013qxa}, may partially explain these CDM small-scale crises, other dark matter models, e.\,g. warm dark matter~\citep{Lovell:2012}, fuzzy dark matter~\citep{Hu:2000ke}, or self-interacting dark matter~\citep{Spergel:1999mh}, have emerged as promising alternatives to CDM. Fuzzy dark matter~(FDM), also known as wave dark matter~\citep{Schive:2014dra, Hui:2021tkt} or ultralight axions~\citep{Marsh:2015xka, Hui:2016ltb, Ferreira:2020fam}, is a particularly interesting DM candidate to resolve the small-scale problems but still maintain CDM predictions on large scales. FDM is typically comprised of extremely light bosons with mass $m_a \sim 10^{-22} \,\eV$, which have de Broglie wavelengths of astronomical scales, $\lambda_{\rm dB} \propto (m_a v)^{-1} \sim \kpc$.

Therefore, in FDM cosmology, bound structures below kiloparsec scales are washed out due to an effective ``quantum'' pressure acting against gravity. For the same reason, FDM haloes form a central flat core called \textit{soliton}~\citep{Schive:2014dra, Schwabe:2016rze, Mocz:2017wlg} rather than the cuspy Navarro-Frenk-White~(NFW) profile of CDM~\citep{Navarro:1995iw}. A typical density profile of FDM haloes also features sophisticated interference granules of density fluctuations~\citep{Schive:2014dra, Mocz:2017wlg} around the central core. However, the simplest model of FDM, where $m_a$ is the only parameter, has been tightly constrained by observational data from the Lyman-$\alpha$ forest~\citep{Armengaud:2017nkf, Irsic:2017yje, Kobayashi:2017jcf}, ultrafaint dwarf galaxies~\citep{Marsh:2018zyw, Hayashi:2021xxu, Dalal:2022rmp}, and strong lensing~\citep{Powell:2023jns}. The current consensus is that a single FDM species of mass $10^{-22} \,\eV$ is unlikely to account for the total dark matter budget while still addressing the small-scale issues of CDM at the same time. 

Several extensions of the most simplified FDM model have been widely explored in literature. A few notable studies are self-interacting FDM~\citep{Mocz:2023adf} where the quartic self-coupling term is included, vector fuzzy dark matter~\citep{Amin:2022pzv} where FDM is a higher-spin field, and mixed cold and fuzzy dark matter~\citep{Schwabe:2020eac}. From a theoretical perspective, the ultralight mass of the FDM field can be generated in the context of string compactifications. It is expected that there may exist more than one species of ultralight axions with masses spanning many orders of magnitudes as in the string axiverse scenario \citep{Svrcek:2006yi, Arvanitaki:2009fg}. Thus, the multiple-field FDM model is naturally motivated by string theory. Along this direction, the multiple-field FDM model with a specific focus on the two-field case has been proposed in \cite{Luu:2018afg}.

We previously found that the ground state of a time-independent two-field FDM system is a nested configuration of two solitons, each from one component field, which is distinct from the usual soliton in the one-field FDM model. Ever since our work, there has been increasing interest in the two-field FDM model, starting with \cite{Eby:2020eas} and \cite{Guo:2020tla} where the authors studied the semi-analytical profile of two-field solitons. More recently, \cite{Huang:2022ffc} performed the first two-field cosmological simulations with CDM initial conditions; \cite{Gosenca:2023yjc} investigated stellar heating in two- and higher-field FDM haloes; \cite{Glennon:2023jsp} simulated soliton collision of two-field FDM with self-interactions and inter-field interactions; \cite{Jain:2023ojg} studied Bose-Einstein condensation in the kinetic regime. These studies showed that the phenomenology of the two-field FDM model can be complex, novel, and compelling. Importantly, it has the potential to retrieve the initial promises of the one-field FDM model.

In this work, we carry out non-cosmological and cosmological simulations of the two-field FDM model to study the final state of virialized haloes. Our main goal is to verify whether the nested soliton, previously predicted in \cite{Luu:2018afg} can emerge and become stable in numerical simulations. The paper is organized as follows. In Sec.~\ref{sec:background}, we review the theoretical background of the two-field \schro--Poisson equations. Sec.~\ref{sec:num_method} describes the numerical method for our simulations and ground-state solver in detail. Sec.~\ref{sec:ideal-simul} and Sec.~\ref{sec:cosmo_simul} set up several simulations and then report the main results and new findings of our paper. Finally, conclusions and potential directions for future work are discussed in Sec.~\ref{sec:conclusion}.

\section{Theoretical Background} \label{sec:background}

In this section, we introduce the time-dependent and time-independent \schro-Poisson equations of the two-field FDM model in Sec.~\ref{sec:two-field-eqs} and Sec.~\ref{sec:scaling-schro-pos}, respectively.

\subsection{Two-field Schrödinger--Poisson equations} \label{sec:two-field-eqs}

In the non-relativistic limit, the minimal (non-self-interacting) system of two-field FDM with different masses is governed by gravitationally coupled \schro--Poisson equations. In physical coordinates in the Newtonian gauge, these equations are given by
\begin{equation}
\begin{aligned}
    &i\hbar\dfrac{\partial\psi_1}{\partial t} + \dfrac{3}{2}H\psi_1 = -\dfrac{\hbar^2}{2m_1} \nabla^2\psi_1 + m_1 \Phi\psi_1, \\
    &i\hbar\dfrac{\partial\psi_2}{\partial t} + \dfrac{3}{2}H\psi_2 = -\dfrac{\hbar^2}{2m_2} \nabla^2\psi_2 + m_2 \Phi\psi_2, \\
    &\nabla^2\Phi = 4\pi G \left(|\psi_1|^2 + |\psi_2|^2 - \bar{\rho} \right),
\end{aligned} \label{Eq:time-depen-schro-poi}
\end{equation}
where $\Phi$ is the gravitational potential, $\psi_1$ and $\psi_2$ are the wavefunctions of the FDM field with masses $m_1$ and $m_2$, respectively. The first field $\psi_1$ is always chosen as the lighter FDM field in our convention throughout this study. The density of each field is given by $\rho_i = |\psi_i|^2$, where $i=1,2$. The total density is obtained by summing individual densities, $\rho = \rho_1 + \rho_2$, while $\bar{\rho}$ is the corresponding mean of the total density. For convenience, we define a parameter to quantify the ratio of the heavy-field mean density to the total density as
 \begin{align}
    \beta_2 \equiv \dfrac{\bar{\rho}_2}{\bar{\rho}_1 + \bar{\rho}_2} = \dfrac{M_1}{M_1 + M_2} \quad {\rm where} \quad \bar{\rho}_i = \dfrac{M_i}{L_{\rm box}^3},
\end{align}
and $M_i$ denotes the total mass of each field enclosed within a volume with length $L_{\rm box}$ along each spatial dimension. As the two-field \schro--Poisson equations \eqref{Eq:time-depen-schro-poi} are highly non-linear due to gravitational couplings, numerical simulations are mandatory to study the system's dynamics. Simulation techniques for the \schro-Poisson equations will be reviewed later in Sec.~\ref{sec:pseudo_spectral}.

\subsection{Time-independent Schrödinger--Poisson equations} \label{sec:scaling-schro-pos}

Some prominent features of FDM haloes, such as the stable ground-state solution, can be found in an approximate limit. Once the field dynamics are stabilized, the time-dependent part of the FDM fields can be factorized out, $\psi_i(\xb,t) = \exp(-iE_it/\hbar) \, \psi_i(\xb)$, where $E_i$ is the ``eigenenergy'' of each field. Let us also assume that the system is spherically symmetric, $\psi_i(\xb) = \psi_i(r), \Phi(\xb) = \Phi(r)$, and omit the Hubble flow as well as the mean density. The time-independent \schro--Poisson equations are then derived as follows
\begin{equation}
\begin{aligned}
    &\dfrac{\partial^2\psi_1}{\partial r^2} = -\dfrac{2}{r} \dfrac{\partial\psi_1}{\partial r} + \dfrac{2m_1}{\hbar^2}(m_1\Phi - E_1)\psi_1, \\
    &\dfrac{\partial^2\psi_2}{\partial r^2} = -\dfrac{2}{r} \dfrac{\partial\psi_2}{\partial r} + \dfrac{2m_2}{\hbar^2}(m_2\Phi - E_2)\psi_2, \\
    &\dfrac{\partial^2\Phi}{\partial r^2} = -\dfrac{2}{r} \dfrac{\partial\Phi}{\partial r} + 4\pi G \left(|\psi_1|^2 + |\psi_2|^2 \right).
\end{aligned} \label{Eq:time-inde-schro-poi}
\end{equation}
The ground-state solution of this system is called a \textit{soliton}, which is a cored density profile that satisfies the following boundary conditions 
\begin{equation}
\begin{aligned}
    &\psi_1(r=0) = \psi_1(0), \quad  \psi_2(r=0) = \psi_2(0), \\
    &\psi_1(r \rightarrow \infty) \rightarrow 0, \quad  \psi_2(r \rightarrow \infty) \rightarrow 0, \\ 
    &\dfrac{\partial\psi_1}{\partial r}(r=0) = \dfrac{\partial\psi_2}{\partial r}(r=0) = 0,
\end{aligned}  
\end{equation}
and $\Phi(0)$ can be arbitrarily chosen. Integrating the density of each field yields the corresponding total mass of the soliton
\begin{align}
    M_{s,i} \equiv \int \rho_i (\xb) d^3\xb = 4\pi \int \rho_i (r) r^2 dr.
\end{align}

In practice, the set of equations \eqref{Eq:time-inde-schro-poi} can be further simplified thanks to the scaling symmetry of the \schro--Poisson equations. The \schro--Poisson equations are unchanged under a transformation of variables with parameter $\lambda$ such that~\citep{Ji:1994xh}
\begin{equation}
    \label{Eq:sp-scaling-relation}
    t \rightarrow \lambda^{-2} t, \quad \xb \rightarrow \lambda^{-1} \xb, \quad \psi_i \rightarrow \lambda^2 \psi_i, \quad \Phi \rightarrow \lambda^2 \Phi.
\end{equation}
This scaling relation suggests that Eqs.~\eqref{Eq:time-inde-schro-poi} can be re-scaled into
\begin{equation}
\begin{aligned}
    &\dfrac{\partial^2\tilde{\psi}_1}{\partial \tilde{r}^2} = -\dfrac{2}{\tilde{r}} \dfrac{\partial\tilde{\psi}_1}{\partial \tilde{r}} + \left(\dfrac{m_1}{m} \right)^2 (\tilde{\Phi} - \tilde{E}_1) \tilde{\psi}_1, \\
    &\dfrac{\partial^2\tilde{\psi}_2}{\partial \tilde{r}^2} = -\dfrac{2}{\tilde{r}} \dfrac{\partial\tilde{\psi}_2}{\partial \tilde{r}} + \left(\dfrac{m_2}{m} \right)^2 (\tilde{\Phi} - \tilde{E}_2) \tilde{\psi}_2, \\
    &\dfrac{\partial^2\tilde{\Phi}}{\partial \tilde{r}^2} = -\dfrac{2}{\tilde{r}} \dfrac{\partial\tilde{\Phi}}{\partial \tilde{r}} + |\tilde{\psi}_1|^2 + |\tilde{\psi}_2|^2,
\end{aligned} \label{Eq:simp-schro-poi}
\end{equation}
where a new set of dimensionless variables has been introduced with a tilde sign. They are related to the original variables by
\begin{equation}
\begin{aligned}
    &r = \dfrac{\hbar^2}{2m^2 GM} \tilde{r}, \quad E_i = \dfrac{2G^2M^2m^2m_i}{\hbar^2} \tilde{E}_i, \\
    &\psi_i = \sqrt{\dfrac{2G^3 M^4 m^6}{\pi \hbar^6}} \tilde{\psi}_i, \quad \Phi = \dfrac{2G^2M^2m^2}{\hbar^2}\tilde{\Phi}.
\end{aligned} \label{Eq:scaling_variables}
\end{equation}
Here, $M$ is a free scaling parameter, corresponding to $\lambda$ in \eqref{Eq:sp-scaling-relation}. $M$ is the normalization factor of the soliton mass, $\int |\tilde{\psi}_i|^2 \tilde{r}^2 d\tilde{r} = M_{s,i}/M$, and $m$ represents an effective FDM mass (which can, however, be freely chosen), which is usually set to $m=\sqrt{m_1m_2}$ in our computations.

\section{Numerical method} \label{sec:num_method}

In this section, we review the pseudo-spectral method used for our simulations in Sec.~\ref{sec:pseudo_spectral}, then provide detailed guidelines to solve for the ground state of two-field equations in Sec.~\ref{sec:ground-state}.

\subsection{Pseudo-spectral solver} \label{sec:pseudo_spectral}

The dynamical two-field FDM system in Eqs.~\eqref{Eq:time-depen-schro-poi} can be simulated using the pseudo-spectral method extended for two fields with different masses. The simulation volume is a cubic box discretized on a fixed Cartesian grid with uniform resolution. Periodic boundary conditions are implicitly imposed in the spectral solver.
The wavefunction of each field is evolved unitarily for each timestep $\Delta t$ as follows:
\begin{itemize}
    \item Step 1: Evolve $\psi_1$ and $\psi_2$ with the potential ``kick'' operator by half a timestep
    \begin{align}
        \psi_i \longleftarrow \exp\left( -i\dfrac{\Delta t}{2}\dfrac{m_i}{\hbar} \Phi \right)\psi_i,
    \end{align}
    \item Step 2: Evolve $\psi_1$ and $\psi_2$ with the kinetic ``drift'' operator by a full timestep in Fourier space
    \begin{align}
        \psi_i \longleftarrow {\rm ifft} \left[  \exp\left( -i\Delta t \dfrac{\hbar}{2m_i}k^2 \right) {\rm fft}[\psi_i] \right],
    \end{align}
    \item Step 3: Calculate the total density and update the potential (also in Fourier space)
    \begin{align}
        \rho = |\psi_1|^2 + |\psi_2|^2, \quad \Phi \longleftarrow {\rm ifft} \left[ -\dfrac{1}{k^2} {\rm fft} \left[ 4\pi G(\rho - \bar{\rho}) \right] \right],
    \end{align}
    \item Step 4: Repeat Step 1 with the updated potential.
\end{itemize}
Here ``fft'' and ``ifft'' represent the Fast Fourier Transform and its inverse. We employ the FDM module of the magneto-hydrodynamics code \textsc{AREPO}~\citep{Springel:2010, Pakmor:2015ana, Weinberger:2019tbd} originally implemented by the authors of \cite{Mocz:2017wlg} and modify it accordingly for two-field FDM simulations.

Since the pseudo-spectral algorithm converges at second order in time and exponentially in space, the heavy-field (smaller) soliton is fully resolved even when it overlaps only a few cells along each dimension. 
On the one hand, the spacing requirement of the grid, $\Delta x$, needs to cover the finest features of the heaviest field~\citep{Mocz:2018ium, May:2021wwp}, i.\,e.,
\begin{align}
    \Delta x < \min \left[ \dfrac{\pi\hbar}{v_{1,\max}m_1}, \dfrac{\pi \hbar}{v_{2,\max}m_2} \right],
\end{align}
where $v_{i,{\rm max}}$ denotes the maximum velocity of the corresponding field. On the other hand, the timestep criterion needs to be satisfied for the lightest field~\citep{Schwabe:2016rze, May:2021wwp}, i.\,e.,
\begin{align}
    \Delta t < \min \left[ \dfrac{4m_1}{3\pi\hbar} \Delta x^2, \dfrac{2\pi\hbar}{m_2|\Phi_{\rm max}|} \right].
\end{align}
These two criteria often require simulations with very high temporal and spatial resolution to resolve the inner core of both fields. Therefore, multiple-field simulations are much more computationally challenging than one-field simulations.

\subsection{Two-field ground state solver} \label{sec:ground-state}

The dimensionless equations \eqref{Eq:simp-schro-poi} are particularly simple to solve and useful to scale up to different physical systems. For instance, in the one-field FDM model, the scaling parameter $m$ can be chosen as the FDM mass $m_a$ and we obtain a completely scale-free set of equations
\begin{align}
    \dfrac{\partial^2\tilde{\psi}}{\partial \tilde{r}^2} = -\dfrac{2}{\tilde{r}} \dfrac{\partial\tilde{\psi}}{\partial \tilde{r}} + (\tilde{\Phi} - \tilde{E}) \tilde{\psi}, \qquad \dfrac{\partial^2\tilde{\Phi}}{\partial \tilde{r}^2} = -\dfrac{2}{\tilde{r}} \dfrac{\partial\tilde{\Phi}}{\partial \tilde{r}} + |\tilde{\psi}|^2, \label{Eq:simp-one-field}
\end{align}
where the field indices have been omitted. As the above equations are independent of FDM parameters, the ground state corresponds to a unique solution of $\tilde{\psi}$. This solution can be freely scaled up (via the parameter $M$), yielding the \textit{soliton profile} approximated by~\citep{Schive:2014dra}
\begin{align}
    |\tilde{\psi}|^2(r) \propto \rho(r) \simeq 1.9 \times 10^9 \dfrac{\Msun}{\kpc^3} \dfrac{ (10^{-22} \,\eV/m_a)^2 ( \kpc/r_{\mathrm{c}} )^4}{ \left[ 1 + 0.091(r/r_{\mathrm{c}})^2 \right]^8},\label{Eq:sol_profile}
\end{align}
where $r_{\mathrm{c}}$ is the core radius defined as the radius where the core density drops by half, $\rho(r_{\mathrm{c}}) = \rho(0)/2$. The total mass of each soliton is given approximately by $M_s \simeq M_{s,0} \, ( r_{\mathrm{c}}/\kpc ) \, ( m_a/10^{-22} \,\eV )^2$ where $M_{s,0} \equiv 2.2 \times 10^8\, \Msun$. We note that this profile only describes the ground-state solution of a one-field FDM system. In contrast, the ground-state solution of a two-field FDM system is not well approximated by this profile. As such, we will refer to the ground state of Eqs.~\eqref{Eq:simp-one-field}, and equivalently the profile \eqref{Eq:sol_profile}, as \textit{one-field soliton}, or simply \textit{soliton}. The ground state of Eqs.~\eqref{Eq:time-inde-schro-poi}, and equivalently Eqs.~\eqref{Eq:simp-schro-poi}, as \textit{two-field soliton} or \textit{nested soliton} whose meaning will become clear later.

For the two-field equations in \eqref{Eq:simp-schro-poi}, we cannot eliminate the FDM masses with scaling relations. Thus, the ground-state solution is dependent on at least two parameters: the FDM mass ratio $m_2/m_1$ and the central wavefunction ratio defined as
\begin{align}
    \alpha_2 \equiv \sqrt{\dfrac{\rho_2(0)}{\rho_1(0)}} = \dfrac{\psi_2(0)}{\psi_1(0)} = \dfrac{\tilde{\psi}_2(0)}{\tilde{\psi}_1(0)}
\end{align}
since $\tilde{\psi}_1(0)$ can always be fixed to a constant, e.\,g., $\tilde{\psi}_1(0) = 1$. We apply the shooting method to solve these equations: (i) pick a not-too-large endpoint $\tilde{r}_{\rm end}$ and arbitrarily choose some initial values for $\tilde{E}_i$; (ii) solve \eqref{Eq:simp-schro-poi} as an initial condition problem where $\tilde{\psi}_1(0) = 1, \tilde{\psi}_2(0) = \alpha_2$; (iii) adjust $\tilde{E}_1$ and $\tilde{E}_2$ simultaneously such that $\tilde{\psi}_i(\tilde{r}_{\rm end}) \rightarrow 0$ with multi-dimensional Newton-Raphson algorithm; (iv) repeat step (i) to (iii) until $\tilde{r}_{\rm end}$ is large enough. There may exist several excited states of different energies satisfying regular boundary conditions. Thus, we must impose the condition $\tilde{\psi}_i(r) > 0~\forall r$ to guarantee the ground-state solution. We list in Tab.~\ref{Tab:eigenenergy} some eigenenergies corresponding to the ground state of the two-field systems for different values of $\alpha_2$ and $\beta_2$ as an example.

\begin{table}
\caption{Dimensionless eigenenergies of two-field solitons for different density and central wavefunction ratios $\beta_2$ and $\alpha_2$ ($m_1 = 8 \times 10^{-23} \,\eV$, $m_2 = 1.6 \times 10^{-22} \,\eV$; same as Scenario~B in Tab.~\ref{Tab:sim_setup}). The first and second entries of each row correspond to $\tilde{E}_1$ and $\tilde{E}_2$, respectively.}

\centering
\begin{tabular}{cccc}
	\hline
	  $\beta_2$ & $\alpha_2$ & $\tilde{E}_i$  (exact) & $\tilde{E}_i$ (approximate) \\
	\hline
	\multirow{2}{*}{0.3} & \multirow{2}{*}{0.68} & 1.4761993842 & 1.4606887253 \\
    & & 0.8739143554 & 0.9756256155 \\
	\hline
    \multirow{2}{*}{0.5} & \multirow{2}{*}{2} & 2.3898610754 & 2.3449988097 \\
    & & 1.5042486006 & 1.5686446461 \\
	\hline
    \multirow{2}{*}{0.7} & \multirow{2}{*}{7.75} & 7.7300571770 &  7.6970072599 \\
    & & 5.0941543699 & 5.1154890489 \\
	\hline
\end{tabular}
\label{Tab:eigenenergy}
\end{table}

Although these steps are straightforward, we cannot apply them to solve the two-field system with a high mass ratio, $m_2/m_1 \gg 1$. In these limits, we need to tune $\tilde{E}_2$ extremely precisely compared to $\tilde{E}_1$ so that the solution does not blow up. As $\tilde{E}_2$ reaches floating-point precision, the shooting method stops converging without arbitrary-precision arithmetic. The high-mass limit is, however, an ideal condition to apply the \textit{flat density approximation}, where the density distribution of the light field is treated as constant with respect to the heavy field assuming that their cores are very different in sizes, $\tilde{r}_{c,2} \ll \tilde{r}_{c,1}$. This allows the use of what are effectively the one-field equations \eqref{Eq:simp-one-field}, with an additional constant external source. The approximation procedure also follows four steps:
\begin{itemize}
    \item Step 1: Solve for the heavy wavefunction with a constant density distribution of the light field as an external source, setting $m = m_1$ and $M = \hat{M}$ (arbitrary), i.\,e.
    \begin{equation}
    \begin{aligned}
        &\dfrac{\partial^2\hat{\psi}_2}{\partial \hat{r}^2} = -\dfrac{2}{\hat{r}} \dfrac{\partial\hat{\psi}_2}{\partial \hat{r}} + (\hat{\Phi} - \hat{E}_2) \hat{\psi}_2 \\
        &\dfrac{\partial^2\hat{\Phi}}{\partial \hat{r}^2} = -\dfrac{2}{\hat{r}} \dfrac{\partial\hat{\Phi}}{\partial \hat{r}} + \left|  \hat{\psi}_2 \right|^2 + \dfrac{1}{\alpha_2^2}
    \end{aligned}
    \;\; {\rm with} \;\; \hat{\psi}_{2}(0) = 1,
    \end{equation}
    where we have adopted an approximation $\hat{\psi}_1 (\hat{r}) \simeq \alpha_2^{-1}$.
    \item Step 2: Rescale the variables by setting $m = m_2$ and $M = \alpha_2^{-1/2} (m_2/m_1)^{3/2} \hat{M}$, resulting in
    \begin{align}
        \psi'_2(r') = \alpha_2 \hat{\psi}_2(\hat{r}), \;\; r' = \alpha_2^{-1/2} \left( m_1/m_2 \right)^{1/2} \hat{r}.
    \end{align}
    \item Step 3: Solve for the light wavefunction with an additional source of the heavy field just found:
    \begin{equation}
    \begin{aligned}
        &\dfrac{\partial^2 \psi'_1}{\partial r'^2} = -\dfrac{2}{r'} \dfrac{\partial\psi'_1}{\partial r'} + (\Phi' - E'_1) \psi'_1 \\
        &\dfrac{\partial^2\Phi'}{\partial r'^2} = -\dfrac{2}{r'} \dfrac{\partial \Phi'}{\partial r'} + |\psi'_1|^2 + \left|  \psi'_2 \right|^2
    \end{aligned}
    \;\; {\rm with} \;\; \psi'_{1}(0) = 1.
    \end{equation}
    \item Step 4: To return to our ``usual choice'' of $m = \sqrt{m_1 m_2}$, rescale the variables again by additionally setting $M = (m_1/m_2)^{3/4} M'$:
    \begin{align}
        \tilde{\psi}_1 (\tilde{r}) = \psi'_1 (r'), \;\; \tilde{\psi}_2 (\tilde{r}) = \psi'_2 (r'), \;\; \tilde{r} = \left( m_2/m_1 \right)^{1/4}  r'.
    \end{align}
    We then obtain an approximate solution of Eqs.~\eqref{Eq:simp-schro-poi}.
\end{itemize}

\begin{figure}
    \centering\includegraphics{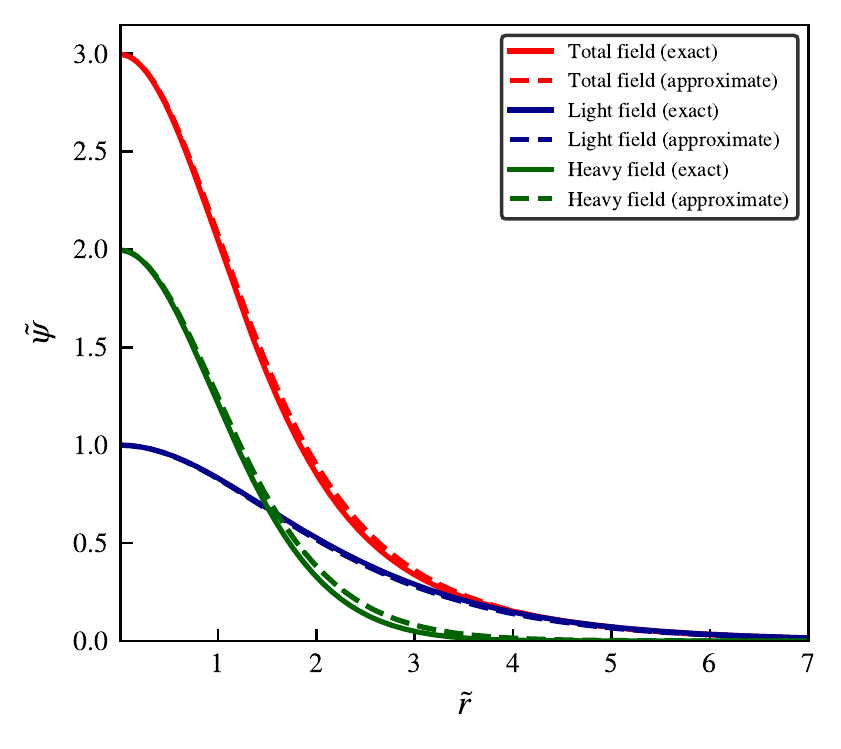}
    \caption{Dimensionless wavefunctions of the ground state with $m_2/m_1 = 2, \beta_2 = 0.5$ and $\alpha_2 = 2$ (see Tab.~\ref{Tab:eigenenergy}) solved by the shooting (solid) and flat density approximation (dashed) methods.}
    \label{fig:wavefunc_test}
\end{figure}

The eigenenergies are not crucial in this method, but they can be straightforwardly retrieved from the scaling relations:
\begin{align}
    E'_2 = \left( m_1/m_2 \right) \alpha_2 \hat{E}_2, \;\; \tilde{E}_i = \left( m_2/m_1 \right)^{1/2} E'_i.
\end{align}
These values provide a reasonable estimate of the exact energies (see Tab.~\ref{Tab:eigenenergy}) and they can serve as an initial guess for the precise shooting method. Figure \ref{fig:wavefunc_test} compares the ground-state wavefunctions solved by both methods for Scenario B with $\beta_2 = 0.5$. Even though the mass ratio is not high in this case, the flat-density approximation (dashed curves) yields virtually indistinguishable profiles from the precise ones (solid curves).

Regardless of the method used, we can restore the two-field soliton density by computing the scaling factor $M$ based on the relation between dimensionless and physical variables. More specifically, we require
\begin{align} 
    M = \left[ \dfrac{\pi \hbar^6 ( 1 + \alpha_2^2 )}{2G^3m_1^3m_2^3\rho(0)} \right]^{1/4}.
\end{align}
to fit a two-field FDM soliton with central density $\rho(0) = |\psi_1(0)|^2 + |\psi_2(0)|^2$. Once $M$ is determined, the physical radius and the physical wavefunction of both fields can be derived according to \eqref{Eq:scaling_variables}.

\begin{figure*}
    \centering
    \hfill \includegraphics{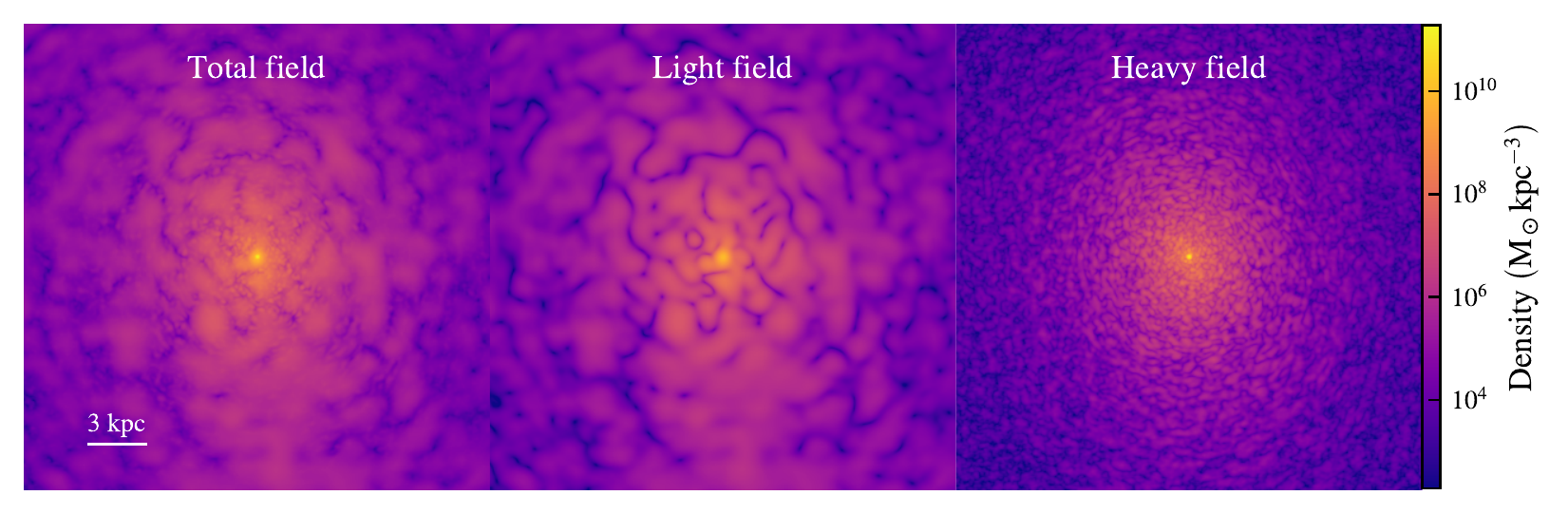} 
    \\
	\includegraphics{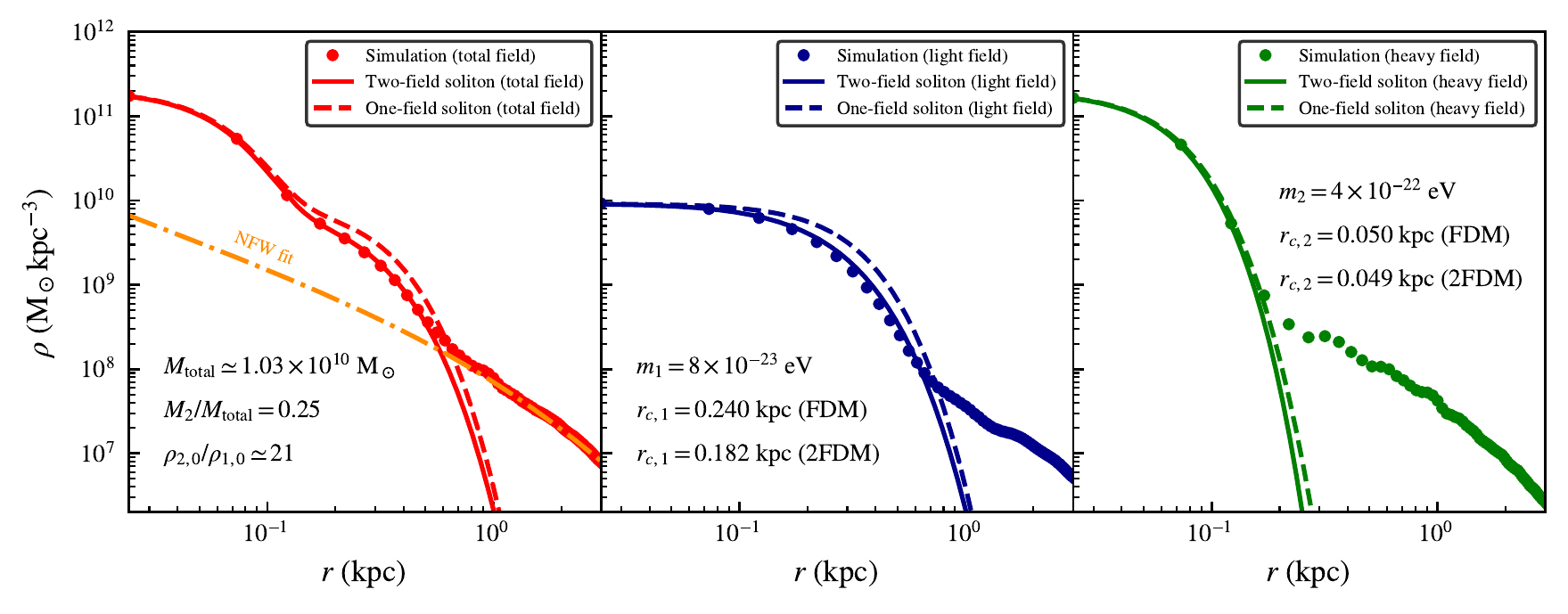}
    \caption{The top panels show slices through a snapshot of the total, light, and heavy fields (from left to right) of a virialized two-field FDM halo at $t \sim 2.6 \,\mathrm{Gyr}$. The bottom panels are the corresponding radial profiles of each field. The red dashed curve in the left panel is shown as the sum of the light (blue dashed) and heavy (green dashed) one-field solitons. The FDM mass of the heavy field is five times that of the light field, $m_2 = 5 m_1 = 4 \times 10^{-22} \,\eV$, while its abundance is four times less than the light field, $\beta_2 = 0.25$, as indicated in the figures. We distinguish the core radius of the best-fit one-field soliton (dashed), denoted as FDM, and two-field soliton (solid), denoted as 2FDM, as the former is overshooting the simulation profile of the light field. In contrast, the core radius of the heavy field is almost identical in both profiles (with dashed and solid green curves almost overlapping). The orange dashed curve denotes the best fit of a NFW profile with the scale density $\rho_{s,0} \sim 9 \times 10^7 \,\Msun \kpc^{-3}$ and the scale radius $r_s \sim 1.85 \,\kpc$ to the outer halo of the total field.}
    \label{fig:col_sim_five}
\end{figure*}

\begin{figure*}
    \centering
    \includegraphics[width=2\columnwidth]{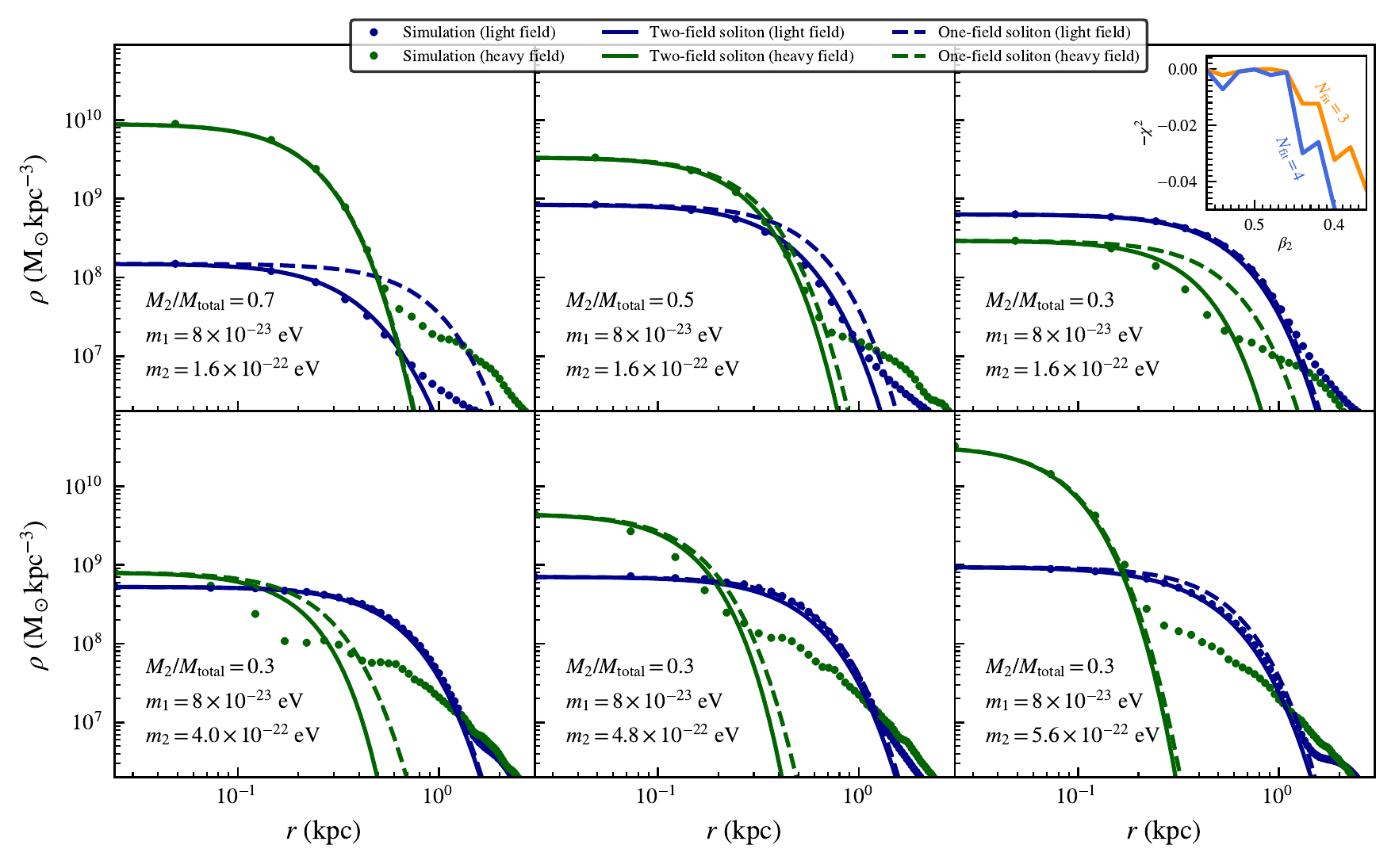}
    \caption{Radial density profiles from several non-cosmological simulations calculated at $t \sim 2.4\,\mathrm{Gyr}$. \textit{(Upper row)} Profiles of the light, $m_1 = 8 \times 10^{-23} \,\eV$, and heavy field, $m_2 = 1.6 \times 10^{-22} \,\eV$, in three simulations with varying mass fractions $\beta_2 \equiv M_2/M_{\rm total}$ (Scenario B). The inset of the last panel shows the goodness-of-fit of the heavy-field profile (see definition in \eqref{eq:chi2}) in other simulations with different $\beta_2$ and at the same moment. \textit{(Lower row)} Profiles of the light and heavy fields in similar simulations with varying heavy-field mass $m_2$ and a fixed mass fraction $\beta_2 = 0.3$ (Scenario C). The heavy field cannot reach a stable ground state in some simulations where neither the two-field soliton nor the one-field soliton fits the simulation profile.}
    \label{fig:col_sim_two}
\end{figure*}

\begin{table}
\caption{Setup of the soliton collision simulations in Sec.~\ref{sec:ideal-simul}. Here the mass of the light field is $m_1 = 8 \times 10^{-23} \,\eV$. Initially, solitons of both FDM species with core radius $r^{\rm(ini)}_{c,i}$ are randomly placed in the simulation box. These solitons are not the same as the two-field soliton, which may (or may not) form as the final state in each scenario. The last column shows whether the two-field soliton is found at the end of each simulation.}

\centering
\begin{tabular}{ccccccc}
	\hline
	  \multirow{2}{*}{Scenario} & \multirow{2}{*}{$\beta_2$} & $M_{\rm total}$ & \multirow{2}{*}{$m_2/m_1$} & $r^{\rm (ini)}_{c,1}$ & $r^{\rm (ini)}_{c,2}$ & Solion \\
     & & $(10^{10} \,\Msun)$ & & (kpc) & (kpc) & formed \\
	\hline
    A & 0.25 & $1.03 $ & 5 & 2 & 0.08 & Yes \\
    \hline
    \multirow{3}{*}{B} & 0.3 & \multirow{3}{*}{$0.3$} & \multirow{3}{*}{2} & 1.64 & 0.95 & Yes \\
     & 0.5 &  &  & 2.29 & 0.57 & Yes \\
     & 0.7 &  &  & 3.82 & 0.41 & No \\
	\hline
    \multirow{3}{*}{C} & \multirow{3}{*}{0.3} & \multirow{3}{*}{$0.3$} & 5 & \multirow{3}{*}{1.64} & 0.15 & No \\
     &  &  & 6 &  & 0.11 & No \\
     &  &  & 7 &  & 0.08 & Yes \\
	\hline
\end{tabular}
\label{Tab:sim_setup}
\end{table}

\section{Soliton collision with two-field FDM} \label{sec:ideal-simul}

We perform non-cosmological simulations to investigate the features of two-field haloes. Initially, the soliton cores with density profiles given by \eqref{Eq:sol_profile} are randomly placed in a box of length $L_{\rm box} = 50 \,\kpc$ with zero velocities. We then let them gravitationally interact, merge, and form a virialized halo. The final state is captured at time $t = 2.4{-}2.6 \,\mathrm{Gyr}$, when the two-field halo has become relatively stationary. Table \ref{Tab:sim_setup} summarizes the setup of the soliton collision simulations in the following sections.

\subsection{Nested soliton profile} \label{sec:nested_soliton}

Firstly, we collide 15 solitons of the light field, $m_1 = 8 \times 10^{-23} \,\eV$, with 45 solitons of the heavy field with $m_2 = 4 \times 10^{-22} \,\eV$ in a simulation box with a resolution of $1024^3$ cells. The core radius of the light-field solitons is manually set to $r_{c,1} = 2 \,\kpc$. The core radius of the heavy-field solitons is inferred from $r_{c,1}$ to match the mass of the light-field ones, $r_{c,2} = r_{c,1} (m_1/m_2)^2$, so that the mean density of the heavy field accounts for 25\% of the total density in the system, i.\,e., $\beta_2 = 0.25$. We refer to this simulation as \textit{Scenario A}.

Figure~\ref{fig:col_sim_five} shows slices of the density distribution and corresponding radial profiles that are averaged over angular directions in Scenario A. The origin is chosen at the location with maximum total density. As expected, the halo of the total field and the individual components show a central core surrounded by interference granules in the simulation. Most notably, the core of the total field exhibits a nested structure of two constituent solitons with a transition around $0.16 \,\kpc$. This nested profile is fitted well with the ground-state soliton (red solid curve) of the time-independent two-field \schro--Poisson equations~\eqref{Eq:time-inde-schro-poi}. Similarly, the predicted two-field soliton profiles of the light field (blue solid curve) and heavy field (green solid curve) are in excellent agreement with the inner cores of the associated simulation fields, although the heavy-field soliton is only resolved by three cells. The outer halo follows an NFW-like profile, similar to the one-field FDM model, as expected.

The one-field profile of the light field (blue dashed curve) with a larger core radius does not fit the simulation data, unlike its two-field counterpart. As this profile is determined by solving the \schro--Poisson equations for the light field only, the corresponding core radius is noticeably wider due to the absence of gravitational attraction from the heavy-field soliton. On the other hand, the one-field and two-field soliton profiles of the heavy field are virtually identical since it only experiences gravitational effects from an almost flat and low-density distribution of the light field, i.\,e., $\rho_2(0) \gg \rho_1(0)$. The one-field profile of the total field (red dashed curve), defined as the sum of two component one-field solitons, is not compatible with the simulation data in a similar manner to the light field.

In summary, the two-field halo profile contains a nested soliton with a visible transition from one core to another and a second transition from the soliton to the remaining NFW-like halo. Yet, the soliton-soliton transition is not always visible due to time-dependent fluctuations of the heavy-field granules which induce a slight density excess over the light-field soliton. The soliton-halo transition is found at $r \sim 3-3.5r_{c,i}$ for the light and heavy field, which agrees with previous studies~\citep{Schive:2014hza, Mocz:2017wlg}. As for the total field, although the soliton-halo transition can be seen around $0.5 \,\kpc$, it is generally ill-defined when the light-field soliton blends with the interference granules of the heavy field as mentioned above.

\subsection{Soliton formation and instabilities} \label{sec:sol_insta}

Even though we have seen the existence of a stable nested soliton as the ground-state solution of two-field \schro--Poisson equations, it may not always be able to form with any value of the mass fraction. Recently, one particular cosmological simulation of two-field FDM~\citep{Huang:2022ffc} with $m_2 = 3m_1 = 3 \times 10^{-22} \,\eV$ and $\beta = 0.25$ has pointed out that the heavy-field soliton cannot form under deep gravitational potential fluctuations induced by the light field. We are, therefore, motivated to study the formation of two-field solitons in \textit{Scenario B} as follows.

We collide 10 solitons of each field with $m_1 = 8 \times 10^{-23} \,\eV$ and $m_2 = 1.6 \times 10^{-22} \,\eV$ in a simulation box with a resolution of $512^3$ cells. The heavy-field solitons are placed concentrically with the light-field ones to enhance collision efficiency. The total mass is fixed, $M_{\rm total} = M_1 + M_2 = 3 \times 10^9 \, \Msun$. The heavy-field density ratio varies from 30\% to 70\%, $\beta_2 =\{ 0.3, 0.5, 0.7 \}$, in three simulations. Thus, the core radii of the initial solitons need to be adjusted in each simulation to satisfy the density ratio required
\begin{equation}
\begin{aligned}
    r_{c,1} &\simeq (1-\beta_2)^{-1}(N_{s,1}M_{s,0}/M_{\rm total}) (10^{-22} \,\eV/m_1)^2, \\
    r_{c,2} &\simeq  \beta_2^{-1}(N_{s,2}M_{s,0}/M_{\rm total}) (10^{-22} \,\eV/m_2)^2, 
\end{aligned} \label{Eq:core_radius}
\end{equation}
where $N_{s,i}$ denotes the number of initial solitons from each FDM species and $N_{s,1} = N_{s,2} = 10$ in this case.

Figure~\ref{fig:col_sim_two} (upper row) shows radial profiles of the light and heavy field in different simulations with $m_2 = 2m_1 = 1.6\times 10^{-22} \,\eV$ and different mass fractions $\beta_2$. We observe distinctive results in each simulation, which are discussed in order as follows. When the heavy field is dominant, $\beta_2 = 0.7$, the two-field solitons fit the numerical simulation well for both component fields. The one-field profile of the heavy field is still aligned with its two-field profile, but a similar alignment is not seen in the case of the light field. These findings completely agree with the previous results in Scenario A. When the two fields are comparable in abundance, $\beta_2 = 0.5$, the halo profile of each field still respects the corresponding two-field ground state. However, both one-field profiles deviate from the two-field ones. When the light field is dominant, $\beta_2 = 0.3$, the inner core of this field is consistent with both one-field and two-field solitons. In contrast, the inner profile of the heavy-field halo fails to fit either of these solitons, which implies that the second soliton cannot form in this case. Our results independently confirm that there exists a limit of a mass fraction where the formation of the heavy-field soliton is inhibited, as pointed out by \cite{Huang:2022ffc}.

We estimate this threshold by running additional simulations which are set up identically as above with $\beta_2$ varying continuously from $0.36$ to $0.54$. We then calculate the goodness-of-fit of the heavy-field soliton which describes how well it fits the corresponding simulation data, defined as
\begin{align}
    \chi^2 = \sum_{i=1}^{N_{\rm fit}} \dfrac{\log\rho_{{\rm sim}, 2}(r_i; \beta_2) - \log\rho_{{\rm sol}, 2}(r_i; \beta_2)}{\log\rho_{{\rm sol}, 2}(r_i; \beta_2)}, \label{eq:chi2}
\end{align}
where $\rho_{{\rm sol}, 2}$ and $\rho_{{\rm sim}, 2}$ are the two-field soliton and the simulation profile of the heavy field, respectively; $N_{\rm fit}$ is the number of simulation data points used counting from the halo center since only these points are expected to fit the soliton. The inset of the upper right panel in Fig.~\ref{fig:cosmo_two_slice_radial} shows $-\chi^2$ with respect to $\beta_2$ for $N_{\rm fit} = 3$ (orange) and $N_{\rm fit} = 4$ (blue). A sudden drop from $\chi^2 \simeq 0$ implies no soliton found in the simulation data. It is, hence, straightforward to infer the mass fraction threshold to be approximately $\beta \simeq 0.44$ for Scenario B.

On the other hand, we must highlight two important properties of two-field FDM haloes that are found in this set of simulations. Firstly, the one-field profiles fail to describe two-field solitons, especially for the light field. These discrepancies are particularly prominent when the heavy field becomes more dominant ($\beta_2 = 0.5$ compared to $\beta_2 = 0.7$ in Scenario A) or when the heavy-field mass is closer to the light-field mass ($m_2=2m_1$ in Scenario B compared to $m_2=5m_1$ in Scenario A). As the heavy-field soliton is more extended and massive in these two limits, its gravitational pull on the light field is also larger and vice versa. The case $\beta_2=0.7$ provides an example when the light soliton is so compressed that its core is no longer flat while the case $\beta_2=0.5$ has the solitons of both fields being squeezed under the gravitational potential of each other. Thus, fitting solely one-field solitons to the numerical simulation data could lead to the non-detection of solitons, which is a misidentification of the real two-field ground state. Secondly, the threshold of a mass fraction where the heavy-field core becomes disrupted is determined by several characteristics of the two-field FDM system. The nested soliton can form with a mass fraction of $\beta_2 = 0.25$ as seen in Scenario A, while the same phenomenon does not happen in the case $\beta_2 = 0.3$ of Scenario B with a higher mass fraction.

\begin{figure*}
    \centering
    \includegraphics{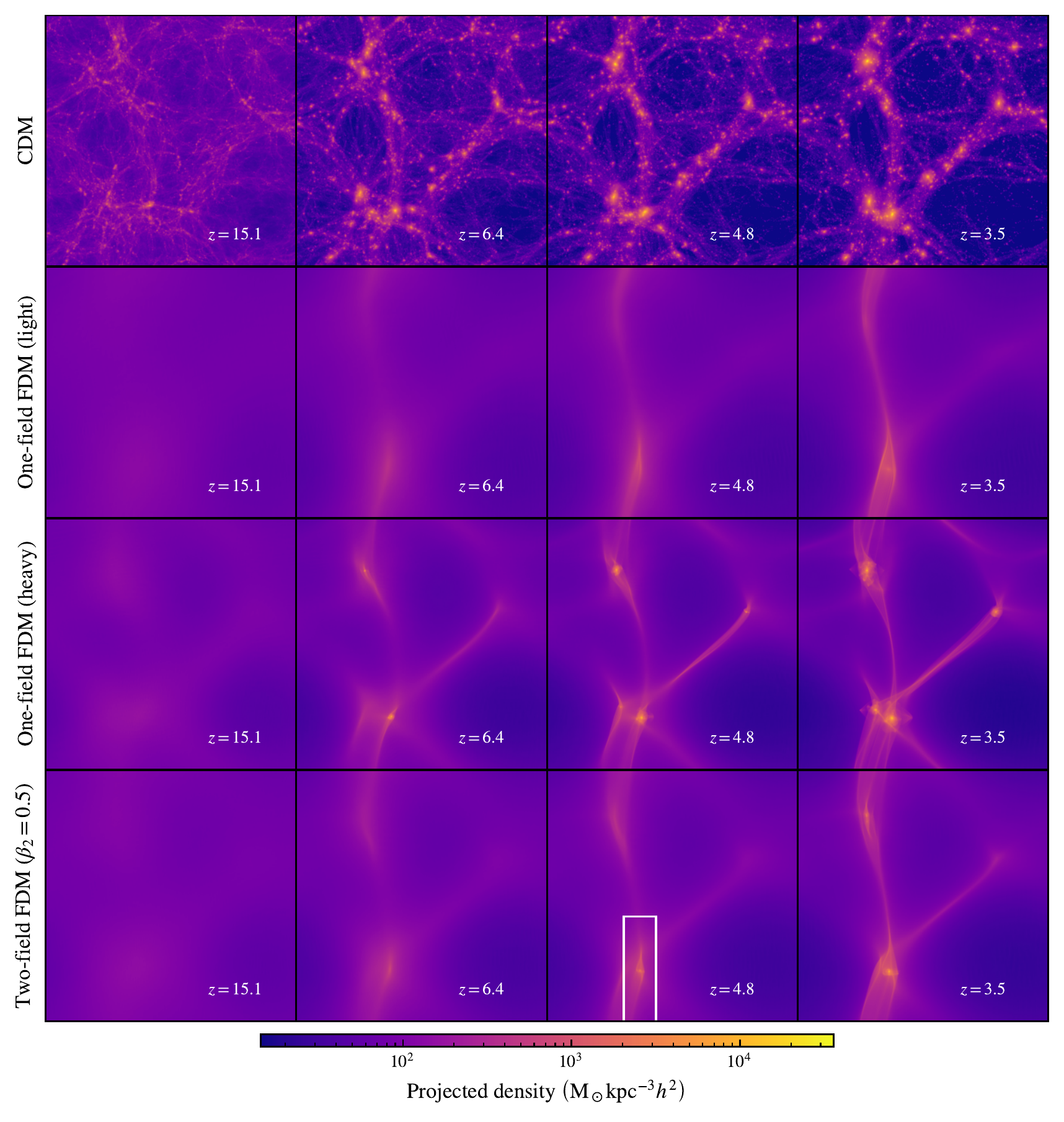}
    \caption{Snapshots of projected (comoving) density in four simulations at $z = 15.1, 6.4, 4.8, 3.5$ to illustrate structure formation of different DM models throughout cosmological history. These values of redshift are chosen to fully capture the onset of soliton formation in each FDM simulation. The simulation volume has a comoving side length of $L_{\rm box} = 1.4 \,h^{-1} \mathrm{Mpc}$. The white box at $z=4.8$ in the bottom row marks the first structure in a two-field FDM cosmology with $m_2 = 2m_1 = 2 \times 10^{-22} \,\eV$ and $\beta_2 = 0.5$. The close-up view of this region is shown in Fig.~\ref{fig:cosmo_two_slice_radial}.}
    \label{fig:cosmo_two_projection}
\end{figure*}

\begin{figure}
    \centering
    \hfill \includegraphics{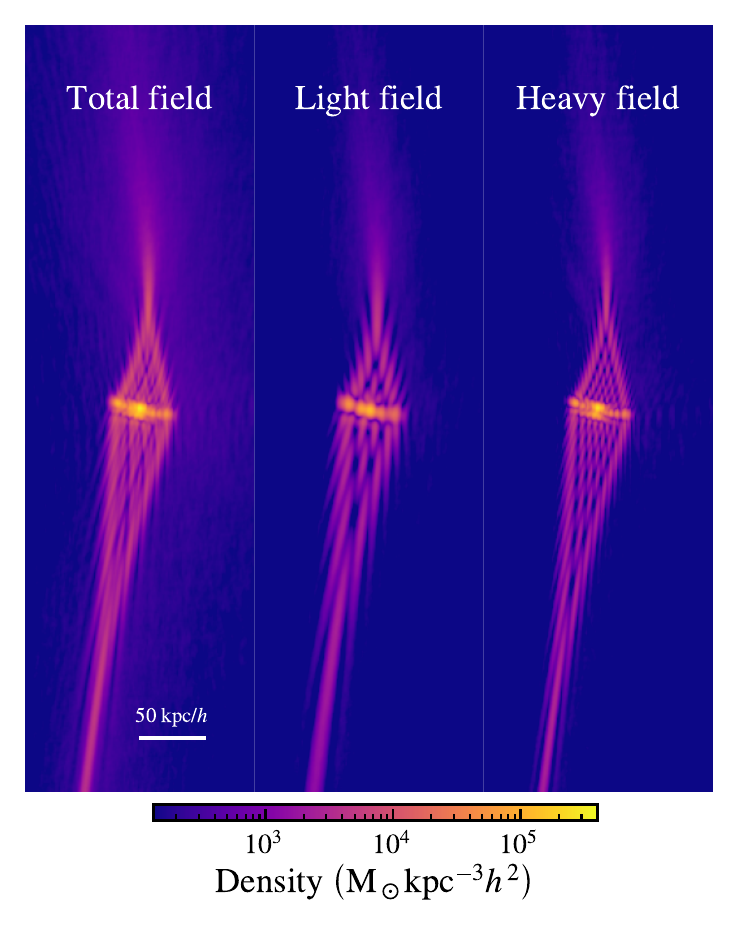} \\
    \includegraphics{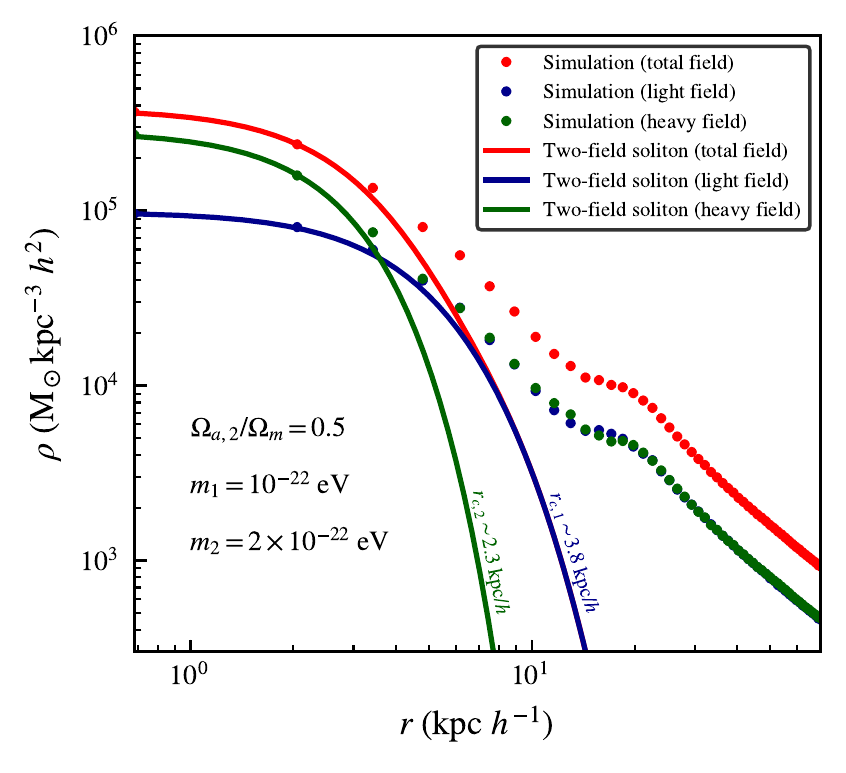}
    \caption{The upper panels show close-up slices through the center of the first two-field halo formed at $z=4.8$ as seen in Fig.~\ref{fig:cosmo_two_projection} (all quantities in comoving coordinates). The lower panel shows the radial profile of this halo in comoving coordinates, including a two-field soliton at the center (red) and an outer NFW-like profile. The profiles of the light field (blue) and the heavy field (green) are also displayed for comparison. Their solitonic cores are just marginally resolved due to the limited resolution of our simulation.}
    \label{fig:cosmo_two_slice_radial}
\end{figure}

However, comparing soliton formation in Scenario A and Scenario B is not completely fair as there are other varying factors, in addition to $\beta_2$ and $m_2$, in these two scenarios such as the total mass $M$ (or equivalently the mean density $\bar{\rho}$) and the initial conditions (core radius, number of colliding solitons). Therefore, we set up \textit{Scenario C} similar to Scenario B but with the heavy-field mass varying, $m_2 = \{ 4.0, 4.8, 5.6 \} \times 10^{-22} \,\eV$, and the density fraction fixed to $\beta_2 = 0.3$ in three simulations. The core radius of initial solitons also changes according to Eq.~\eqref{Eq:core_radius}.

The final profiles are shown in the lower row of Fig.~\ref{fig:col_sim_two}, where the heavy-field soliton can only form with $m_2 = 7m_1 = 5.6 \times 10^{-22} \,\eV$, but not with $m_2 = 5m_1 = 4 \times 10^{-22} \,\eV$ and $m_2 = 6m_1 = 4.8 \times 10^{-22} \,\eV$, for a fixed $\beta_2 = 0.3$. The only difference in the last case with $m_2=7m_1$ is that its initial state at $t=0$ consists of much denser solitons (see Eq.~\ref{Eq:sol_profile} and Tab.~\ref{Tab:sim_setup}) which enables the system to reach gravitational equilibrium faster. Once the stable soliton of the heavy field has formed, it is no longer susceptible to tidal disruption from density fluctuations of the light field. Thus, it is likely that the formation of a nested soliton in the two-field FDM model is viable if the heavy-field soliton formation has completed relatively earlier than the light-field one. This condition can be realized even if the heavy-field density is sub-dominant as long as the two FDM masses differ by orders of magnitude, $m_2 \gg m_1$.

\section{Two-field FDM in cosmology} \label{sec:cosmo_simul}

We perform DM-only cosmological simulations of the two-field FDM model with self-consistent initial conditions.
 
\subsection{Simulation setup} \label{sec:cosmo_setup}

Four simulations are conducted in a comoving volume that has a length of $L_{\rm box} = 1.4 \,h^{-1} \mathrm{Mpc}$ with a spatial resolution of $1024^3$ cells. The four simulations are set up as follows: (i) CDM via $N$-body dynamics described by Vlasov-Poisson equations, (ii) one-field FDM with $m_a = m_1 = 10^{-22}~\eV$ for the light field only (i.\,e.\ $\beta_2 = 0$), (iii) one-field FDM with $m_a = m_2 = 2 \times 10^{-22}~\eV$ for the heavy field only (i.\,e.\ $\beta_2 = 1$), (iv) two-field FDM with $m_1$ and $m_2$ as given above and $\beta_2 = 0.5$. We note that the mass fraction of two fields in this case is defined as the ratio of cosmological density parameters, i.\,e.,
\begin{align}
    \beta_2 = \Omega_{\mathrm{a},2}/\Omega_{\mathrm{m}} = \Omega_{\mathrm{a},2}/(\Omega_{\mathrm{a},1}+\Omega_{\mathrm{a},2}),
\end{align}
where $\Omega_{\mathrm{a},i}$ denotes the present-day density fraction of each FDM field.
 
Initial conditions are generated from the same random seed for comparison at $z=127$ using \textsc{N-GenIC} from~\cite{Volker:2015} with a two-field matter power spectrum calculated by the publicly available Boltzmann code \textsc{aHCAMB} from~\cite{Luu:2021yhl}\footnote{The code \textsc{aHCAMB}, based on \textsc{CAMB}~\citep{Lewis:1999bs} and \textsc{axionCAMB}~\citep{Grin:2022}, was originally written for the axi-Higgs model. Thus, we modify it to include additional perturbative equations of two axion fields indirectly coupled via the metric tensor.}. The heavy-field mass is assumed to be $m_2 = 2 \times 10^{-22} \,\eV$, two times heavier than the light-field mass of $m_1 = 10^{-22} \,\eV$. Cosmological parameters are chosen as constrained by Planck~\citep{Planck:2018vyg}, i.\,e., $\Omega_{\mathrm{m}} = 0.314, \Omega_{\Lambda} = 0.686, h = 0.6732, n_s = 0.966$. One exception is the primordial spectrum amplitude, which is boosted to $A_s = 7.5 \times 10^9$ to enhance small-scale fluctuations. Thus, the simulation box is more representative of an overdense region than an average cosmological volume in the two-field FDM universe. The physical system in these simulations evolved until redshift $z \simeq 3.5$, after which halo cores become too small in the comoving grid to resolve.

\subsection{Results and Discussion} \label{sec:cosmo_result}

Figure \ref{fig:cosmo_two_projection} provides a glimpse of the evolution of DM clustering in cosmological simulations of four different DM models, as set up in Sec.~\ref{sec:cosmo_setup}. Each panel shows the projected density distribution of DM structures which consists of various DM haloes connected by cosmic filaments. As expected from previous FDM simulations~\citep{Mocz:2019pyf, May:2022gus}, small-scale fluctuations are smoothed out in all FDM models, compared to the scale-independent clustering in the CDM model. The number of FDM haloes is, thus, strongly suppressed due to a sharp cut-off below the Jeans scale in the initial matter power spectra and the ``quantum pressure'' from FDM dynamics that counteracts gravity. In terms of one-field simulations, the heavy-field structures are more abundant and form earlier than the light-field ones throughout cosmic history. The first structure of the heavy field emerges at $z \simeq 6.4$, while the first light-field halo is not seen until $z \simeq 3.5$. We also identify four heavy-field haloes in total compared to only one light-field halo at the final redshift.

On the other hand, the two-field FDM cosmology (with $\beta_2=0.5$) looks like an intermediate state of the one-field models. Two haloes have formed by the end of our simulation with the first one becoming visible at $z \simeq 4.8$. Zooming into this two-field halo in Fig.~\ref{fig:cosmo_two_slice_radial} (upper panel) reveals the soliton core and wave interference patterns on scales of the de Broglie wavelength of both fields. Since the total field is a superposition of two-component fields, its core and granular structures are slightly denser and fuzzier. We found that the radial profile of the central soliton exactly follows the two-field ground state (lower panel), as in the non-cosmological simulations, even though the solitons are just barely resolved. The nested structure is not apparent in this case, owing to the low mass ratio of two fields, i.\,e., $m_2 = 2m_1$.

More importantly, the haloes in our two-field simulation consist of solitons of both light and heavy fields. These findings show that the light-field soliton formation in the two-field simulation is greatly enhanced, as no soliton appeared until the last redshift in the only-light-field simulation. The light-field density likely started clustering faster thanks to deep gravitational potentials induced by overdense regions of heavy-field haloes which had formed beforehand in the two-field cosmology. Therefore, in the limit of high mass ratios, $m_2 \gg m_1$, we speculate that heavy-field soliton formation may not be susceptible to the oscillating potential of the light-field soliton in a realistic cosmological context, even with low-density fractions $\beta_2$ between two fields, as its clustering commences significantly earlier in the cosmic history. On the other hand, late-time DM haloes may predominantly incorporate solitons of both fields in this context, which poses a new challenge to explain the diverse core size of dwarf galaxies by multiple FDM species~\citep{Luu:2018afg, Pozo:2023zmx}, where soliton formation of each FDM species should be mostly independent.

\section{Concluding remarks} \label{sec:conclusion}

As evidence has begun to build up against the one-field FDM model with mass $m_a \sim 10^{-22} \,\eV$, the two-field FDM model with different masses $m_1$ and $m_2$ offers an interesting and plausible alternative for the FDM paradigm. In this study, we have conducted state-of-the-art simulations to study structure formation in the two-field FDM model. With simulations of soliton collision, we discovered that the virialized haloes feature a nested soliton and an NFW-like outer profile. The nested soliton is a stable configuration of two concentric cores of different sizes, which can be solved as the ground state of two-field time-independent \schro--Poisson equations.

Unlike the one-field soliton profile, component solitons in the two-field model are more compact due to mutual gravitational attraction. Depending on the density ratio of the component fields, soliton formation may not occur in case the heavy field is excessively dominated by the light field. Deep gravitational potentials induced by soliton oscillations may be responsible for this phenomenon~\citep{Huang:2022ffc}. However, this threshold of soliton instabilities is also sensitive to initial conditions and mass differences between the two fields. A general pattern is that as the second field becomes heavier, the nested soliton can form at a lower density fraction.

We also had a preliminary look into the large-scale structure of the two-field FDM model. Despite the limited resolution, two-field soliton profiles are found in our cosmological simulation. Since there are three viable profiles in two-field FDM cosmology: two-field nested soliton, one-field light soliton, and one-field heavy soliton, one might expect to find these different types of haloes in a larger volume, which would explain the separate hierarchy of core sizes in spheroidal dwarf galaxies and ultrafaint dwarf galaxies~\citep{Luu:2018afg}. Unfortunately, every halo in our small-box simulation only contains a two-field soliton, so we leave this interesting problem to a future study. 

The two-field model can potentially alleviate several tensions between FDM predictions and observational data. \cite{Hayashi:2021xxu} argued that ultrafaint dwarf galaxy cores as FDM solitons imply $m_a \gtrsim 10^{-21} \,\eV$, incompatible with the $10^{-22} \,\eV$ mass for dwarf spheroidal galaxies, which can be precisely explained by the heavy field. \cite{Dalal:2022rmp} suggested that orbiting stars are perturbed as they interact with the gravitational potential of density granules in FDM haloes, which is particularly prominent for lighter FDM fields, so the mass range $m_a \lesssim 3 \times 10^{-19} \,\eV$ is strongly excluded. As indicated by \cite{Gosenca:2023yjc}, the presence of one (or more) heavier field(s) helps to smooth out the density fluctuations in FDM haloes, hence relaxing the bound. Lastly, FDM suppresses the linear matter power spectrum at small scales, which alters the structure of the intergalactic medium~(IGM) on the scales probed by the Lyman-$\alpha$ forest of early-time quasars. By relating the one-dimensional flux spectrum with the FDM power spectrum, \cite{Armengaud:2017nkf} and \cite{Irsic:2017yje} constrained $m_a \lesssim \times 10^{-21} \,\eV$. \cite{Kobayashi:2017jcf} also derived similar constraints for the mixed DM model for any amount of FDM more than $30\%$ of the total DM. Although there still exist several uncertainties of IGM gas physics assumed in these studies, which should be better understood~\citep{Hui:2016ltb}, we expect a sufficiently large density fraction of the heavy field will resolve the Lyman-$\alpha$ constraint. Nevertheless, more detailed studies on the two-field FDM model in theory, phenomenology, and simulation are needed to settle these issues.

Although our work provides a fairly comprehensive study of two-field FDM regime, the results are restricted to the low mass ratio $m_2/m_1=2-7$ and the minimal model with uncoupled fields. It is, therefore, interesting to explore different variations of the two-field FDM in a broader context. For instances, \cite{Schwabe:2020eac} have extensively studied the mixed DM model and pointed out that the soliton cannot form in cosmological simulations with FDM constituting less than $10\%$ of the total DM. Provided CDM is effectively treated as the heavy FDM field with much higher mass than the light field, $m_2 \gg m_1$, there seems to exist a density threshold that inhibits the soliton formation of the light field, which deserves more investigation. Another example is \cite{Cyncynates:2021xzw} who proposed a coupled ``friendly'' two-axion model with a cosine potential. As long as the heavy-field mass is sufficiently close to the light-field one, $m_2/m_1 \sim 0.75$, the coupling triggers autoresonance mechanism which synchronizes the background oscillations of both fields. Furthermore, small-scale density fluctuations are enhanced in this model, which can alleviate the Lyman-$\alpha$ constraints mentioned above, a feature shared by the extreme axion framework with an initial large displacement angle~\citep{Zhang:2017dpp, Arvanitaki:2019rax}.

At the end of the day, since the nested soliton is exclusive to two (or multiple) species of FDM, potential observations of this profile in future galaxy surveys will be a smoking gun signature of the multiple-axion scenario.

\section*{Acknowledgements}

The simulations in this paper were run on the Engaging cluster supported by MIT.
H.N. thanks Mihir Kulkarni for help with setting up the initial conditions of two-field cosmological simulations, Xuejian Shen, and Leo Fung for useful discussions to improve the paper.
H.N., T.B. and H.T. acknowledge support from the Research Grants Council of Hong Kong through the Collaborative Research Fund C6017-20G.
P.M. acknowledges this work was in part performed under the auspices of the U.S. Department of Energy by Lawrence Livermore National Laboratory under contract DE-AC52-07NA27344, Lawrence Livermore National Security, LLC.
S.M. acknowledges support from the National Science Foundation under Grant No.\ 2108931.

\section*{Data availability}
The data underlying this article will be shared on reasonable request to the corresponding author.

%%%%%%%%%%%%%%%%%%%%%%%%%%%%%%%%%%%%%%%%%%%%%%%%%%
%%%%%%%%%%%%%%%%%%%% REFERENCES %%%%%%%%%%%%%%%%%%

\bibliographystyle{mnras}
\bibliography{references}

%%%%%%%%%%%%%%%%%%%%%%%%%%%%%%%%%%%%%%%%%%%%%%%%%%

\bsp
\label{lastpage}
\end{document}